\documentclass[letterpaper, 10 pt, conference]{ieeeconf}
\IEEEoverridecommandlockouts
\usepackage{caption}      
\usepackage{amsmath,amssymb,amsfonts}
\usepackage{graphicx}
\usepackage{textcomp}
\usepackage{xcolor}
\def\BibTeX{{\rm B\kern-.05em{\sc i\kern-.025em b}\kern-.08em    T\kern-.1667em\lower.7ex\hbox{E}\kern-.125emX}}
\usepackage{verbatim}
\usepackage{paralist}
\usepackage{graphics} 
\usepackage{epsfig} 
\usepackage{mathptmx} 
\usepackage{times} 
\usepackage{physics}
\usepackage{mdwmath}
\usepackage{mdwtab}
\usepackage[hidelinks]{hyperref}
\usepackage{algpseudocode}
\usepackage{algorithm}
\usepackage{caption}
\usepackage{subcaption}
\usepackage{cite}
\usepackage{algorithm}    
\usepackage{algpseudocode} 
\usepackage{balance}

\usepackage{tikz}
\usetikzlibrary{shapes, arrows.meta, positioning}

\newtheorem{proposition}{Proposition}%
\newtheorem{definition}{Definition}
\newtheorem{remark}{Remark}

\def\BibTeX{{\rm B\kern-.05em{\sc i\kern-.025em b}\kern-.08em
    T\kern-.1667em\lower.7ex\hbox{E}\kern-.125emX}}
\begin{document}

\title{Quantum Solution Framework for Finite-Horizon LQG Control via Block Encodings and QSVT\\

\thanks{The authors acknowledge the support of the Danish e-Infrastructure Consortium (DeiC) and the National Quantum Algorithm Academy (NQAA) through the Postdoctoral Scholarship 
under the project ``Quantum-Driven Solutions for Multi-Agent Systems and Advanced Computation''. This work was also partially supported by UID/00147- Research Center for Systems and Technologies (SYSTEC) - and the Associate Laboratory Advanced Production and Intelligent Systems (ARISE, 10.54499/LA/P/0112/2020) funded by Fundação para a Ciência e a Tecnologia, I.P./ MCTES through the national funds.}
}

 \author{Nahid Binandeh Dehaghani, Rafal Wisniewski, A. Pedro Aguiar
 \thanks{N. Dehaghani and R. Wisniewski are with Department of Electronic Systems, Aalborg University, Fredrik Bajers vej 7c, DK-9220 Aalborg, Denmark
 {\tt\small nahidbd@es.aau.dk, raf@es.aau.dk}}
\thanks{A. Aguiar is with the Research Center for Systems and Technologies (SYSTEC), Electrical and Computer Engineering Department, FEUP - Faculty of Engineering, University of Porto, Rua Dr. Roberto Frias sn, i219, 4200-465 Porto, Portugal
        {\tt\small pedro.aguiar@fe.up.pt}}%
 }

\maketitle

\begin{abstract}
We present a quantum algorithm for solving the finite-horizon discrete-time Linear Quadratic Gaussian (LQG) control problem, which integrates optimal control and state estimation in the presence of stochastic disturbances and noise. Classical approaches to LQG require solving a backward Riccati recursion and a forward Kalman filter, both requiring computationally expensive matrix operations with overall time complexity $\mathcal{O}(T n^3)$, where $n$ is the system dimension and $T$ is the time horizon. 
While efficient classical solvers exist, especially for small to medium-sized systems, their computational complexity grows rapidly with system dimension. To address this, we reformulate the full LQG pipeline using quantum linear algebra primitives, including block-encoded matrix representations and quantum singular value transformation (QSVT) techniques for matrix inversion and multiplication. We formally analyze the time complexity of each algorithmic component. Under standard assumptions on matrix condition numbers and encoding precision, the total runtime of the quantum LQG algorithm scales polylogarithmically with the system dimension $n$ and linearly with the time horizon $T$, offering an asymptotic quantum speedup over classical methods. 
\end{abstract}

\section{Introduction} 
Linear Quadratic Gaussian (LQG) control is one of the most celebrated results in modern control theory, integrating optimal feedback control and state estimation in the presence of stochastic noise. It offers a powerful solution framework for systems governed by linear dynamics and subjected to Gaussian process and measurement uncertainties, combining the Linear Quadratic Regulator (LQR) and the Kalman filter via the separation principle \cite{anderson2007optimal, whittle1990risk,simon2006optimal, athans1971role, shaiju2008formulas}.

Solving the finite-horizon discrete-time LQG problem classically involves solving a backward Riccati recursion and a forward Kalman filter—both requiring dense matrix operations such as multiplications and inversions. These steps incur a computational complexity of $\mathcal{O}(n^3)$ per time step, making LQG computationally intensive for large-scale, high-dimensional systems. With the growing demand for scalable and real-time optimal control, there is increasing interest in exploring alternative computational paradigms that can overcome these classical bottlenecks.

Quantum computing has emerged as a promising frontier in computational mathematics, particularly for linear algebra problems. Quantum linear systems algorithms (QLSAs), such as the Harrow-Hassidim-Lloyd (HHL)  \cite{harrow2009quantum}, and more general frameworks like Quantum Singular Value Transformation (QSVT) \cite{gilyen2019quantum}, offer the potential to perform matrix inversion and other operations exponentially faster than classical methods under suitable conditions. Central to these techniques is the notion of block encoding, which embeds classical matrices into unitary operators, enabling efficient quantum implementation of arithmetic operations on large matrices \cite{lin2022lecture,sunderhauf2024block,clader2022quantum,chakraborty2018power}.

In this work, we propose a quantum algorithm for solving the finite-horizon discrete-time LQG control problem. Our approach restructures the classical LQG pipeline into a fully quantum framework.
The backward Riccati recursion (for control gain synthesis) and the forward Kalman filter (for optimal state estimation) are both implemented using quantum subroutines based on block-encoded matrix arithmetic and QSVT-based matrix inversion. The proposed algorithm provides a unified quantum framework that delivers control inputs and estimated states with 
logarithmic complexity in the system dimension, assuming availability of efficient block encodings, well-conditioned matrices, and access to quantum-accessible data structures.

This contribution builds on and extends recent efforts in quantum control and quantum filtering \cite{shi2024quantum,clayton2024differentiable,krovi2024quantum}, and connects the LQG problem to recent advances in quantum scientific computing \cite{lin2022lecture,gilyen2019quantum,chakraborty2018power,camps2022fable}. Beyond the immediate application to control synthesis and estimation, our framework also establishes a foundation for future research into online quantum filtering, and scalable optimal control of quantum and classical dynamical systems. \\

Our main contributions are as follows:
\begin{itemize}
    \item {Quantum Reformulation of LQG:} We propose a fully quantum algorithm that solves the finite-horizon discrete-time LQG control problem, integrating both optimal control synthesis and state estimation in a unified framework.
    
    \item {Block-Encoding-Based Implementation:} We reformulate the Riccati recursion and Kalman filtering steps using block-encoded matrix arithmetic and QSVT, enabling efficient quantum execution of these control-theoretic procedures.
    
    \item {Complexity Analysis:} We provide a detailed analysis of the quantum time complexity, showing that under standard assumptions (efficient data access, well-conditioned matrices), the quantum algorithm achieves polylogarithmic scaling in the system dimension and linear scaling in the time horizon, improving upon the classical $\mathcal{O}(Tn^3)$ cost.\\
    
\end{itemize}

\paragraph*{Assumptions and Scope}
This work establishes a theoretical framework for quantum solution methods in finite-horizon LQG control. We assume access to fault-tolerant quantum hardware capable of executing block-encoded linear algebra routines and QSVT-based transformations with bounded error, along with idealized data access through quantum-accessible data structures that enable efficient state preparation and matrix encoding. In particular, we assume that any matrix \( A \in \mathbb{C}^{n \times n} \) used in a quantum subroutine is accessed via a block encoding with a known normalization factor \( \alpha \geq \|A\| \). For simplicity, we often take \( \alpha = 1 \) in algorithmic descriptions and complexity analysis. However, in general, \( \alpha \) may scale with the spectral or Frobenius norm of the matrix, which can grow with the system dimension depending on the problem structure. Since the runtime of QSVT-based algorithms typically depends linearly on \( \alpha \), this factor can significantly influence the overall complexity. \\

The remainder of this paper is structured as the following: Section II reviews the classical formulation of the finite-horizon LQG control problem, presenting both the state feedback and output feedback cases. Section III introduces the quantum algorithmic primitives required for our approach, including block encoding techniques, quantum-accessible data structures, and Quantum Singular Value Transformation (QSVT). In Section IV, we present the proposed quantum algorithm for LQG control, detailing both the backward Riccati recursion for optimal control synthesis and the forward Kalman filtering for state estimation within the block-encoding framework. Section V provides a complexity analysis of the quantum algorithm, comparing its theoretical performance with classical methods.

\section{Classical Finite-Horizon Linear Quadratic Gaussian (LQG) Control}

In this section, we consider the discrete-time finite-horizon LQG control problem. The system is governed by linear dynamics subject to additive Gaussian noise, and the control objective is to minimize a quadratic cost functional over a finite time horizon. Depending on whether the system state is fully or partially accessible, the solution structure changes accordingly. We begin with the case where the full system state is available to the controller.
\subsection{State Feedback Case}
We consider a discrete-time stochastic control problem over a finite horizon \( k = 0, 1, \dots, T - 1 \). The system state at time \( k \) is denoted by \( x_k \in \mathbb{R}^n \), and the control input by \( u_k \in \mathbb{R}^m \).
The system dynamics are governed by the linear stochastic difference equation:
\begin{equation}\label{9}
    x_{k+1} = A x_k + B u_k + w_k, \quad k = 0, 1, \dots, T - 1,
\end{equation}
where the process noise sequence \( \{w_k\} \) consists of independent and identically distributed (i.i.d.) Gaussian random vectors with zero mean and covariance \( \Sigma \), i.e., \( w_k \sim \mathcal{N}(0, \Sigma) \). The process is defined on a probability space \( (\Omega, \mathcal{F}, \mathbb{P}) \), and expectations are taken with respect to the measure \( \mathbb{P} \). It is further assumed that \( x_k \) and \( w_k \) are independent for all \( k \).

The control objective is to minimize the expected value of a quadratic cost functional over the horizon:
\begin{equation}\label{10}
    J(u) := \mathbb{E} \left[ \sum_{k=0}^{T-1} c(x_k, u_k) + x_T^T M_T x_T \right],
\end{equation}
where \( c(x_k, u_k) \) denotes the stage cost,
\[
c(x, u) := x^T M x + u^T N u + 2 x^T S u,
\]
and \(M_T \ge 0 \) is the terminal cost matrix. Here, we consider that the covariances and weighting matrices are time-invariant, for simplicity of presentation.

If the initial state \( x_0 \) is Gaussian \( x_0 \sim \mathcal{N}(\mu_0, R_0) \),  
it follows from the linear-Gaussian nature of the system that each \( x_k \) is also Gaussian with \( x_k \sim \mathcal{N}(\mu_k, R_k) \), where the mean and covariance of the state evolve according to:
\begin{equation}
    \mu_{k+1} = A \mu_k + B u_k, \quad R_{k+1} = \Sigma + A R_k A^T.
\end{equation}
We now present the following proposition on state feedback LQG control, based on established results in
\cite{shaiju2008formulas, whittle1990risk}.\\
\begin{proposition}
Consider the stochastic system defined by \eqref{9} and the cost functional in \eqref{10}. Assume that \[
\begin{bmatrix}
M & S \\
S^T & N
\end{bmatrix} \geq 0, \quad M_T \geq 0, \quad N > 0.
\] holds, then the optimal control policy that minimizes the expected cost is given by:
\begin{equation}
    u_k^* = K_k x_k, \quad k = 0, 1, \dots, T - 1
\end{equation}
where \( K_k \) is the state feedback gain obtained from the standard discrete-time Riccati recursion,
\begin{equation}\label{kk}
 K_k := -\left( N + B^T P_{k+1} B \right)^{-1} \left( S^T + B^T P_{k+1} A \right).   
\end{equation}
Furthermore, the corresponding value function has the form:
\begin{equation}
    \bar{F}_k(x) = x^T P_k x + r_k,
\end{equation}
where \( P_k \) is the solution to the backward Riccati recursion 
\begin{equation}\label{pk}
    \begin{aligned}
      P_k &= M + A^T P_{k+1} A - \left( S^T + B^T P_{k+1} A \right)^T 
\left( N + B^T P_{k+1} B \right)^{-1}\\
&\left( S^T + B^T P_{k+1} A \right)  
    \end{aligned}
\end{equation}
with terminal condition \( P_T = M_T \), and \( r_T = 0 \). The scalar term \( r_k \) is computed as:
\begin{equation}
    r_k = \sum_{j = k+1}^{T} \text{Tr}(\Sigma P_j), \quad k = 0, 1, \dots, T - 1.
\end{equation}
\end{proposition}

\subsection{Output Feedback Case}
We now consider the case where the system state is not fully accessible. The dynamics of the system remain as described in \eqref{9}, with the initial state assumed to be Gaussian \( x_0 \sim \mathcal{N}(\mu_0, R_0) \). The noisy measurements \( y_k \in \mathbb{R}^p \) are given by:
\begin{equation}\label{y}
    y_{k+1} = C x_k + v_{k},
\end{equation}
where \( v_k \sim \mathcal{N}(0, \Gamma) \) denotes measurement noise. The process and observation noise may be correlated, with the joint covariance matrix of \( [w_k, v_k]^T \) denoted by
$ \Delta =
    \begin{bmatrix}
        \Sigma & \Upsilon \\
        \Upsilon^T & \Gamma
    \end{bmatrix}$,
under the assumption \( \Gamma> 0 \).

The cost function remains the same as defined in \eqref{10}. In this setting, let \( \mu_k = \mathbb{E}[x_k \mid y_1, \dots, y_k, u_0, \dots, u_{k-1}] \) denote the conditional mean of the state given the observation and control histories. The corresponding error covariance is denoted by \( R_k \), and we define the information state as \( \chi_k = (\mu_k, R_k) \).

The estimate \( \mu_k \) evolves according to the Kalman filter recursion:
\begin{equation}
    \mu_{k+1} = A \mu_k + B u_k + L_{k+1} ( y_{k+1} - C \mu_k ),
\end{equation}
where the Kalman gain \( L_k \) is given by:
\begin{equation}
    L_{k+1} = (\Upsilon + A R_k C^T) ( \Gamma + C R_k C^T )^{-1}.
\end{equation}

The Riccati recursion governing the evolution of the covariance matrices \( R_k \) is:
\begin{align}\label{13}
    R_{k+1} &= \Sigma + A R_k A^T \notag \\
    &\quad - (\Upsilon + A R_k C^T)(\Gamma + C R_k C^T)^{-1}(\Upsilon + A R_k C^T)^T.
\end{align}
Building on the foundational results in \cite{shaiju2008formulas, whittle1990risk}, we recall the following proposition concerning output feedback LQG control.\\

\begin{proposition}
Consider the stochastic system with partial observations defined by \eqref{9}, \eqref{y}, and the cost functional in \eqref{10}:

\begin{itemize}
    \item[(i)] The optimal control minimizing the expected cost is:
    \[
        u_k^* = K_k \mu_k,
    \]
    where \( K_k \) is the same state feedback gain obtained from the Riccati equation used in the fully observable case.

    \item[(ii)] The corresponding value function is a function of the information state \( \chi_k = (\mu_k, R_k) \), and is given by:
    \[
        \bar{F}_k(\chi_k) = \mu_k^T P_k \mu_k + s_k,
    \]
    where \( P_k \) is the solution to the backward Riccati recursion, and the scalar term \( s_k \) is computed via:
    \[
        s_k = \text{Tr}(R_k M) + \sum_{j=k}^{T-1} \text{Tr} \left[ R_j M + (\Sigma + A R_j A^T - R_{j+1}) P_{j+1} \right].
    \] \\
\end{itemize}
\end{proposition}

\begin{remark}
The Riccati recursion for the error covariance matrix \( R_k \) in \eqref{13} can be compactly expressed as:
\[
    R_{k+1} = p(R_k),
\]
where the mapping \( p(\cdot) \) is defined by:
\[
    p(R) = \Sigma + A R A^T - (\Upsilon + A R C^T)(\Gamma + C R C^T)^{-1}(\Upsilon + A R C^T)^T.
\]

Alternative equivalent expressions for \( p(R) \) and the Kalman gain \( L_{k+1} \) are given by:
\begin{equation*}
  \begin{aligned}
       p(R) =& \Sigma - \Upsilon \Gamma^{-1} \Upsilon^T + (A - \Upsilon \Gamma^{-1} C) 
    (C^T \Gamma^{-1} C + R^{-1})^{-1} \\    
   & (A - \Upsilon \Gamma^{-1} C)^T, 
\end{aligned}  
\end{equation*}
\begin{equation*}
    L_{k+1} = \Upsilon \Gamma^{-1} + (A - \Upsilon \Gamma^{-1} C) 
    (C^T \Gamma^{-1} C + R^{-1})^{-1} C^T \Gamma^{-1}.
\end{equation*}
This alternate formulation remains valid even when \( R \) is singular. In such cases, one can replace the inverse term \( (C^T \Gamma^{-1} C + R^{-1})^{-1} \) with \( (R C^T \Gamma^{-1} C + I)^{-1} R \). \\
\end{remark}

Having established the classical formulation and solution of the LQG problem, we now explore how Algorithm \ref{alg:classical-LQR} can be restructured in the context of quantum algorithms to potentially yield computational advantages for large-scale systems.

\begin{algorithm}[t] 
\caption{Finite-Horizon Output Feedback LQG Controller}
\label{alg:classical-LQR}
\begin{algorithmic}[1]
    \Statex \textbf{Inputs:}
    \State Horizon length \( T \)
    \State System matrices \( A, B, C \)
    \State Cost matrices \( M, N, S \), terminal cost $M_T$
    \State Noise covariances \( \Sigma, \Gamma \), cross-covariance \( \Upsilon \)
    \State Initial estimate \( \mu_0 \), initial covariance \( R_0 \)

    \Statex \textbf{Backward Pass: Solve Riccati Equation for LQG Control}
    \State Set \( P_T \gets M_T \)
    \For{\( k = T-1 \) down to \( 0 \)}
        \State Compute optimal feedback gain:
        \[
        K_k \gets -\left( N + B^T P_{k+1} B \right)^{-1} \left( S^T + B^T P_{k+1} A \right)
        \]
        \State Update Riccati matrix:
\begin{equation*}
    \begin{aligned}
          P_k \gets & M + A^T P_{k+1} A - (S^T + B^T P_{k+1} A)^T \left( N + B^T P_{k+1} B \right)^{-1}\\ &(S^T + B^T P_{k+1} A)      
    \end{aligned}
\end{equation*}

    \EndFor

    \Statex \textbf{Forward Pass: Kalman Estimation and Optimal Control}
    \State Compute initial control: \( u_0 \gets K_0 \mu_0 \)

    \For{\( k = 0 \) to \( T-1 \)}
        \State \textbf{(1) Receive measurement: observe} \( y_{k+1} \)
        \State Kalman gain:
        \[
        L_{k+1} \gets (\Upsilon + A R_k C^T)(\Gamma + C R_k C^T)^{-1}
        \]
        \State Update state estimate:
        \[
        \mu_{k+1} \gets A \mu_k + B u_k + L_{k+1}(y_{k+1} - C \mu_k)
        \]
        \State Update error covariance:
        \begin{equation*}
            \begin{aligned}
               R_{k+1} \gets& \Sigma + A R_k A^T - \\
               &(\Upsilon + A R_k C^T)(\Gamma + C R_k C^T)^{-1}(\Upsilon + A R_k C^T)^T 
            \end{aligned}
        \end{equation*}

        \State \textbf{(2) Compute control input:} \( u_{k+1} \gets K_{k+1} \mu_{k+1} \)
    \EndFor

    \Statex \textbf{Outputs:}
    \State Optimal control inputs \( \{ u_0, \dots, u_{T-1} \} \)
    \State Estimated states \( \{ \mu_1, \dots, \mu_T \} \)
\end{algorithmic}
\end{algorithm}

\section{Quantum Primitives for Block-Encoding-Based LQG Solution}
Classically, solving the finite-horizon discrete-time Linear Quadratic Gaussian (LQG) control problem is computationally intensive. It involves repeated matrix operations such as multiplications, additions, and inversions, resulting in a computational complexity that scales cubically with the system dimension.

This computational burden motivates the exploration of quantum algorithms that offer exponential or polynomial speedups via quantum linear algebra subroutines. In particular, quantum algorithms can potentially offer efficient solutions to large-scale LQG problems by leveraging the principles of block encoding and amplitude amplification.

To reformulate the LQG problem in a quantum computational framework, we adopt the technique of quantum block encoding, which embeds classical matrices into higher-dimensional unitary operators. This embedding enables quantum circuits to perform linear algebra operations such as matrix multiplication, inversion, and linear system solving with polylogarithmic complexity overhead, assuming access to efficient quantum data structures.

Block encoding is the foundation of recent QLSAs, which we aim to use for both components of the LQG pipeline. Specifically, we use block-encoded matrix arithmetic and QSVT to implement the Riccati and Kalman recursions. 
In what follows, we review the key quantum primitives required to implement this framework, with emphasis on block encoding, quantum linear system solving, and efficient quantum representations of structured matrices relevent to control problems.

\subsection{Block Encoding of Classical Matrices}

Block encoding is a core technique in quantum computing that enables classical matrices to be embedded into larger unitary matrices, enabling their use within quantum circuits for efficient arithmetic operations. The central idea is to represent a matrix \( A \in \mathbb{C}^{n \times n} \) as the top-left block of a unitary operator \( U_A \), scaled by a normalization factor \( \alpha > 0 \). Formally, if \( A \) acts on \( s \) qubits, it can be encoded into a unitary acting on \( s + a \) qubits as:
\[
U_A =
\begin{bmatrix}
A/\alpha & * \\
* & *
\end{bmatrix},
\]
where \( * \) denotes unspecified subblocks and \( U_A \in \mathbb{C}^{2^{s+a} \times 2^{s+a}} \).

Let \( |\psi\rangle \) be an arbitrary quantum state on \( s \) qubits, and let \( |0^a\rangle \) denote \( a \) ancilla qubits initialized in the zero state. The embedded matrix \( A \) can be retrieved by preparing the joint state \( |0^a\rangle \otimes |\psi\rangle \), applying the unitary \( U_A \), and post-selecting on the ancilla qubits being in state \( |0^a\rangle \). The result of this procedure is:
\[
A = \alpha \left( \langle 0^a | \otimes I_s \right) U_A \left( |0^a\rangle \otimes I_s \right),
\]
which extracts the normalized top-left block of \( U_A \).

Following the formulations in \cite{sunderhauf2024block, clader2022quantum, gilyen2019quantum, lin2022lecture}, we formalize the definition of a block encoding:

\begin{definition}[Block Encoding]
Let \( A \in \mathbb{C}^{n \times n} \) act on \( s \) qubits. Suppose there exist a normalization factor \( \alpha > 0 \), error tolerance \( \epsilon > 0 \), ancilla count \( a \in \mathbb{N} \), and a unitary matrix \( U_A \in \mathbb{C}^{2^{s+a} \times 2^{s+a}} \), such that
\[
\left\| A - \alpha \left( \langle 0^a | \otimes I_s \right) U_A \left( |0^a \rangle \otimes I_s \right) \right\| \leq \epsilon,
\]
then \( U_A \) is called an \( (\alpha, a, \epsilon) \)-block encoding of \( A \). That is, \( A \) appears approximately in the top-left block of \( U_A \), up to normalization and bounded error \( \epsilon \).\\
\end{definition}

\begin{definition}[Block Encoding Set]
Following \cite{lin2022lecture}, we denote the set of all block encodings of \( A \) with given parameters by \( \mathrm{BE}_{\alpha, a}(A, \epsilon) \). The subset of exact block encodings (i.e., with zero error) is denoted by \( \mathrm{BE}_{\alpha, a}(A) := \mathrm{BE}_{\alpha, a}(A, 0) \).\\
\end{definition}

\begin{remark}[Matrix Padding for Block Encoding]
If \( A \in \mathbb{C}^{m \times n} \) with \( m,n \leq 2^s \) for some integer \( s \), then \( A \) can be embedded into a larger square matrix \( A_e \in \mathbb{C}^{2^s \times 2^s} \), where the top-left \( m \times n \) block contains \( A \), and the remaining entries are set to zero. The padded matrix \( A_e \) can then be used to construct a block encoding with respect to the full \( 2^s \)-dimensional Hilbert space \cite{gilyen2019quantum}.
\end{remark}

\subsection{Constructing Block Encodings from Quantum Data Structures}

To apply block encoding in practice, one must construct a unitary \( U_A \) that embeds a given matrix \( A \) into a larger quantum circuit. Several approaches exist for this construction, many of which rely on quantum-accessible data structures that allow efficient preparation of matrix rows or columns as quantum states \cite{lin2022lecture,camps2024explicit,camps2022fable,clader2022quantum}.

A widely used method builds the block encoding \( U_A \) as a composition of two unitaries, \( U_L \) and \( U_R \), which prepare quantum states proportional to the rows and columns of the matrix \( A \), respectively. The following result, adapted from \cite{chakraborty2018power}, formalizes this construction.\\

\begin{proposition}[Block Encoding via Quantum Data Access]
Let \( A = [A_{ij}] \in \mathbb{C}^{m \times n} \), where \( m, n \leq 2^s \) for some integer \( s \), which corresponds to the number of qubits required to index the rows and columns of \( A \).
Suppose \( A \) is stored in a quantum-accessible data structure that allows coherent access to its entries. Then, there exist unitaries \( U_L \) and \( U_R \), which can be implemented in time
$\mathcal{O}\left(\mathrm{polylog}\left(mn/\epsilon\right)\right)$,
such that the composition
$U_A = U_L^\dagger U_R$
defines an \((\|A\|_F, s, \epsilon)\)-block encoding of matrix \( A \), where \( \|A\|_F \) is the Frobenius norm of \( A \), defined as
$\|A\|_F = \sqrt{\sum_{i,j} |A_{ij}|^2}$.
The unitaries \( U_L \) and \( U_R \) are given by:
\begin{align*}
U_L: \ket{0^s} \ket{j} &\mapsto \frac{1}{\|A\|_F} \sum_{i=1}^{m} \|A_{i,\cdot}\| \ket{i} \ket{j}, \\
U_R: \ket{i} \ket{0^s} &\mapsto \ket{i} \left( \frac{1}{\|A_{i,\cdot}\|} \sum_{j=1}^{n} A_{ij} \ket{j} \right), 
\end{align*}
where \( \|A_{i,\cdot}\| \) is the \(\ell_2\)-norm of the \( i \)-th row of \( A \), and the ancilla register \( \ket{0^s} \) is used to construct the full unitary \( U_A  \).
\end{proposition}


\subsection{Arithmetic Operations on Block Encodings}

Once a matrix has been embedded via block encoding, it is possible to perform arithmetic operations such as addition and multiplication directly in the encoded form. These operations are enabled by the Linear Combination of Unitaries (LCU) framework and are foundational to quantum linear algebra algorithms. The results presented here are adapted from \cite{gilyen2019quantum,lin2022lecture} and tailored to the structure of our LQG formulation.\\

\begin{proposition}[Addition of Block-Encoded Matrices]
Let  \( U_A \) be an \( (\alpha, a, \delta) \)-block encoding of an $s$-qubit operator \( A \),
and \( U_B \) be a \( (\beta, b, \epsilon) \)-block encoding of an $s$-qubit operator \( B \).
Then, there exists a \( (\alpha + \beta, \max(a,b) + 1, \delta + \epsilon) \)-block encoding \( W \) of \( A + B \), constructed as follows:

\begin{enumerate}
    \item {Linear Combination:} \\
    Define the operator \( T = \alpha U_A \oplus \beta U_B \), where \( \oplus \) denotes zero-padding to ensure \( T \) acts on \( \max(a, b) \) ancilla qubits. The top-left block of \( T \) satisfies the bound:
    \begin{align*}
&\left\| (A + B) 
- \left( \bra{0}^{\otimes \max(a,b)} \otimes I_s \right) 
T 
\left( \ket{0}^{\otimes \max(a,b)} \otimes I_s \right) 
\right\| 
 \notag \\
& \leq \delta + \epsilon.
\end{align*}

    \item {Unitary Embedding:} 
    Let \( U = \ket{0}\bra{0} \otimes U_A + \ket{1}\bra{1} \otimes U_B \), and let \( V \) be a unitary matrix such that
    $V = \frac{1}{\sqrt{\alpha + \beta}} 
\begin{bmatrix}
\sqrt{\alpha} & * \\
\sqrt{\beta} & *
\end{bmatrix}$.
    Then, the operator \( W = (V^\dagger \otimes I_s) U (V \otimes I_s) \) is a \( (\alpha + \beta, \max(a,b) + 1) \)-block encoding of \( T \), with top-left block \( T / (\alpha + \beta) \).

    \item {Error Bound:} The approximation error of \( W \) as a block encoding of \( A + B \) is bounded by \( \delta + \epsilon \).\\
\end{enumerate}

\end{proposition}

\begin{proposition}[Multiplication of Block-Encoded Matrices]
Let \( U_A \) be an \( (\alpha, a, \delta) \)-block encoding of an \( s \)-qubit operator \( A \), and \( U_B \) be a \( (\beta, b, \epsilon) \)-block encoding of an \( s \)-qubit operator \( B \). Then the product unitary
$U := (I_b \otimes U_A)(I_a \otimes U_B)$
is an \( (\alpha\beta, a + b, \alpha \epsilon + \beta \delta) \)-block encoding of the product \( AB \).
\end{proposition}


\subsection{Matrix Inversion via QSVT}

Matrix inversion is a fundamental subroutine in many quantum algorithms, including those for solving linear systems and control problems. In this work, we leverage Quantum Singular Value Transformation (QSVT)~\cite{gilyen2019quantum, martyn2021grand} to enable efficient matrix inversion in a block-encoded setting.

Consider a square matrix \( A \in \mathbb{C}^{n \times n} \) with singular value decomposition (SVD) \( A = W \Sigma V^\dagger \), where \( \Sigma = \mathrm{diag}(\sigma_1, \dots, \sigma_n) \) is the diagonal matrix of singular values. We assume that the singular values lie within the interval \( [\kappa^{-1}, 1] \), where \( \kappa \) is the condition number of \( A \). The inverse of \( A \) is then formally expressed as
\[
A^{-1} = V \Sigma^{-1} W^\dagger = V f(\Sigma) W^\dagger,
\]
where \( f(x) = x^{-1} \). 
In the QSVT framework, this odd function is approximated by applying a bounded polynomial \( p(x) \approx x^{-1} \) with odd parity to the singular values via a sequence of quantum operations on a block-encoding of \( A \). The polynomial \( p(x) \) is constructed such that
\[
\left| p(x) - \frac{1}{\kappa \beta x} \right| \leq \epsilon', \quad \forall x \in \left[-1, -\frac{1}{\kappa}\right] \cup \left[\frac{1}{\kappa}, 1\right],
\]
for some normalization constant \( \beta \in \mathbb{R}_+ \), and approximation error \( \epsilon' > 0 \). The odd parity ensures Hermiticity when the polynomial is applied to a Hermitian-embedded matrix, and the constraint \( |p(x)| \leq 1 \) for \(x \in \left[-1, -\frac{1}{\kappa}\right] \cup \left[\frac{1}{\kappa}, 1\right] \) ensures boundedness over the full spectral domain.

QSVT guarantees that a unitary \( U_{p(A)} \in \mathrm{BE}_{1, a+1}(p(A)) \) can be constructed, which is a block-encoding of the matrix \( p(A) \approx A^{-1} \), with precision \( \epsilon' \). The degree of the polynomial scales as \( \mathcal{O}(\kappa \log(1/\epsilon')) \). This block-encoded inverse plays an important role in our quantum algorithm for computing the Riccati-based feedback gains and Kalman filter updates in the finite-horizon LQG control setting.

\section{Quantum Algorithm for Finite-Horizon LQG Control}

The quantum implementation of the finite-horizon output-feedback LQG controller evolves quantum-encoded matrices and vectors through sequences of unitary operations. Classical control inputs and state estimates are extracted via measurement at the end of the procedure. This approach unifies optimal control and state estimation—via Riccati recursion and Kalman filtering—executed entirely on block-encoded representations of system and cost matrices.

Before presenting the complete quantum procedure, we clarify two key assumptions that underpin the algorithm design and complexity analysis:

\begin{itemize}
    \item {Problem Setup:} All system matrices—namely the dynamics matrices 
    \( A \in \mathbb{R}^{n \times n} \), \( B \in \mathbb{R}^{n \times m} \); 
    cost matrices 
    \( M \in \mathbb{R}^{n \times n} \), \( N \in \mathbb{R}^{m \times m} \); 
    cross-term 
    \( S \in \mathbb{R}^{n \times m} \); 
    and noise covariances 
    \( \Sigma \in \mathbb{R}^{n \times n} \), \( \Gamma \in \mathbb{R}^{p \times p} \), and 
    \( \Upsilon \in \mathbb{R}^{n \times p} \)—are assumed to be constant and known in advance. The initial information state \( (\mu_0, R_0) \) is also given, and the 
    measurements \( y_k \in \mathbb{R}^p \) are available at every time step \( k \).
    
    \item {Block-Encoding Assumption:} 
    For complexity analysis, we assume that any matrix \( A \in \mathbb{C}^{n \times n} \) used in the algorithm can be stored in a quantum-accessible data structure and block-encoded into a unitary using 
    \( \mathcal{O}(\mathrm{polylog}(n/\epsilon)) \) resources. For simplicity, we further assume that all matrices in the algorithm are either exactly block-encoded or approximated with error
    $\epsilon$ small enough that it does not affect the asymptotic complexity.
\end{itemize}

The quantum LQG procedure consists of two main stages:
\begin{itemize}
    \item {Backward Pass:} Quantum Riccati recursion for computing optimal state feedback gains.
    \item {Forward Pass:} Quantum Kalman filtering for optimal state estimation and computation of the control input \( u_k \) at each time step.
\end{itemize}
Both stages involve matrix addition, multiplication, and inversion, implemented via quantum subroutines applied to block-encoded matrices. In what follows, we detail each stage of the quantum LQG algorithm, beginning with the backward Riccati recursion.

\subsection{Quantum Backward Pass: LQR via Block-Encoded Riccati Recursion}
We now present a quantum algorithm to compute the LQR feedback gains \( K_k \) by solving the Riccati recursion backward in time using block-encoded quantum linear algebra. The algorithm assumes access to block encodings of the system matrices \( A, B \) and cost matrices \( M, N, S, M_T \).

At each time step \( k \), the gain matrix \( K_k \) and Riccati matrix \( P_k \) are computed based on \eqref{kk} and \eqref{pk}.
These operations are implemented using the following quantum primitives:
\begin{itemize}
    \item {Block Encodings}: All matrices are represented as \( (\alpha, a, \epsilon) \)-block encodings with efficient quantum data access.
    \item {Matrix Multiplication and Addition}: Implemented via LCU-based arithmetic on block-encoded matrices.
    \item {Matrix Inversion}: Performed using QSVT with a polynomial approximation of \( f(x) = x^{-1} \).
\end{itemize}

The full backward pass procedure is summarized in Algorithm~\ref{alg:quantum-riccati}.

\begin{algorithm}[t] 
\caption{Quantum Riccati Recursion for LQR Control}
\label{alg:quantum-riccati}
\begin{algorithmic}[1]
\Statex \textbf{Input:}
\State Time horizon \( T \)
\State Block encodings \( U_A, U_B, U_M, U_{M_T}, U_N, U_S \)
\State QSVT polynomial parameters for matrix inversion
\Statex \textbf{Initialization:}
\State Set \( U_{P_T} \gets U_{M_T} \)

\For{\( k = T-1 \) down to \( 0 \)}
    \State Compute block encoding: \( U_1 \gets U_B^T \cdot U_{P_{k+1}} \cdot U_B + U_N \)
    \State Compute inverse: \( U_{1^{-1}} \gets \text{QSVT-Invert}(U_1) \)   
    \State Compute intermediate matrix:

   $ U_2 \gets U_S^T + U_B^T \cdot U_{P_{k+1}} \cdot U_A$
    \State Compute gain matrix:
    $U_{K_k} \gets - U_{1^{-1}} \cdot U_2$
    \State Update Riccati matrix:

    $U_{P_k} \gets U_M + U_A^T \cdot U_{P_{k+1}} \cdot U_A  - U_2^T \cdot U_{1^{-1}} \cdot U_2$
\EndFor

\Statex \textbf{Output:}
\State Block-encoded matrices \( \{ U_{K_0}, \dots, U_{K_{T-1}} \} \) and \( \{ U_{P_0}, \dots, U_{P_T} \} \)
\end{algorithmic}
\end{algorithm}


\subsection{Quantum Forward Pass: Kalman Filtering via Block Encodings}

We now present the quantum implementation of the Kalman filter, which recursively estimates the system state based on noisy measurements and prior control inputs. This forward pass produces the information state \( (\mu_k, R_k) \), where \( \mu_k \) is the conditional mean of the state given past observations, and \( R_k \) is the error covariance. The construction follows the formulation and quantum algorithm proposed in \cite{shi2024quantum}, adapted here for integration into the LQG control pipeline.

\paragraph{Block-Encoding Setup}
In our quantum implementation, we assume that all system matrices—namely \( A, B, C, \Sigma, \Gamma, \Upsilon \)—as well as time-varying matrices such as the error covariance \( R_k \), innovation covariance \( V_k \), and Kalman gain \( L_k \), admit efficient block encodings. These block encodings allow matrix operations—including additions, multiplications, and QSVT-based inversion—to be implemented efficiently within the Kalman filtering recursion.


\paragraph{Kalman Gain Construction}
The quantum computation of the Kalman gain \(L_k \) is carried out by constructing block encodings of the matrices:
\[
W_k := \Upsilon + A R_k C^T, \quad V_k := \Gamma + C R_k C^T.
\]
These matrices are computed via block-encoded matrix additions and multiplications, implemented via the LCU framework. 

The innovation covariance matrix \( V_k \) is symmetric and assumed to be well-conditioned. To compute a block encoding of its inverse, we apply the QSVT technique using a polynomial approximation of the inverse function. The resulting unitary implements a block encoding of \( V_k^{-1} \), which is then multiplied with the block encoding of \( W_k \) to obtain the block-encoded Kalman gain:
$L_k = W_k V_k^{-1}$.
This final matrix multiplication is implemented using the block-encoded matrix product rule. Under standard assumptions, this process yields a valid quantum representation of the Kalman gain for use in the forward pass.


\paragraph{Covariance Update.}
The update of the error covariance matrix \( R_{k+1} \) is carried out using a sequence of block-encoded matrix operations:
\[
R_{k+1} = \Sigma + A R_k A^T - L_k V_k L_k^T.
\]
Each term in this expression is implemented using block-encoded matrix multiplication and addition, leveraging the linear combination of unitaries (LCU) technique for quantum arithmetic.

We assume that all input matrices, including \( A, \Sigma, L_k, V_k \), and the recursively updated \( R_k \), are available as \( (\alpha, a, \epsilon) \)-block encodings. The matrix product \( A R_k A^T \) is computed by composing two block-encoded multiplications, which results in a new block encoding with updated normalization \( \alpha^2 \), ancilla width \( 2a \), and total error \( \mathcal{O}(\alpha \epsilon) \). Similarly, the innovation term \( L_k V_k L_k^T \) is computed via two sequential multiplications
$U_1 := L_k V_k, \quad U_2 := U_1 L_k^T$,
each of which increases the normalization and ancilla count multiplicatively and additively, respectively.
Intermediate block encodings are rescaled as needed, and amplitude amplification can be applied when necessary to maintain accuracy. The resulting block-encoded matrix \( R_{k+1} \) is then used in the next iteration of the Kalman filter. This recursive structure ensures compatibility with the block-encoding framework and supports efficient quantum propagation of the error covariance over the full time horizon.

We summarize the full forward pass of the quantum Kalman filter in Algorithm~\ref{alg:quantum-kalman}. This algorithm enables recursive estimation of the system state by propagating block-encoded representations of the conditional mean \( \mu_k \) and error covariance \( R_k \) across time. Each iteration involves matrix multiplications, additions, and one QSVT-based inversion, all performed with logarithmic ancilla overhead and bounded approximation error. When combined with the backward Riccati recursion for computing the LQR feedback gains \( K_k \), the quantum Kalman filter completes the forward pass of the finite-horizon LQG controller. 

\begin{algorithm}[t] 
\caption{Quantum Kalman Filter via Block Encodings}
\label{alg:quantum-kalman}
\begin{algorithmic}[1]
\Statex \textbf{Input:}
\State Time horizon \( T \)
\State Block-encoded initial state \( U_{\mu_0} \), initial covariance \( U_{R_0} \)
\State Block encodings \( U_A, U_B, U_C, U_\Sigma, U_\Gamma, U_\Upsilon \)
\State Block encodings of quantum measurement vectors \( \{ U_{y_1}, \dots, U_{y_T} \} \)
\Statex \textbf{Initialization:}
\State Prepare block-encoded inputs \( U_{\mu_0}, U_{R_0} \)

\For{\( k = 0 \) to \( T-1 \)}
    \State Construct innovation covariance:
    $U_{V_k} \gets U_\Gamma + U_C \cdot U_{R_k} \cdot U_C^T$
      \State Invert via QSVT:
    $U_{V_k^{-1}} \gets \text{QSVT-Invert}(U_{V_k})$
    \State Construct intermediate numerator:
    $U_{W_k} \gets U_\Upsilon + U_A \cdot U_{R_k} \cdot U_C^T$
       \State Kalman gain (block encoding):
    $U_{L_{k+1}} \gets U_{W_k} \cdot U_{V_k^{-1}}$
         \State Compute innovation vector:
    $U_{\delta_{k+1}} \gets U_{y_{k+1}} - U_C \cdot U_{\mu_k}$   
    \State Update mean estimate:

    $U_{\mu_{k+1}} \gets U_A \cdot U_{\mu_k} + U_B \cdot U_{u_k} + U_{L_{k+1}} \cdot U_{\delta_{k+1}}$    
    \State Update error covariance:

$    U_{R_{k+1}} \gets U_\Sigma + U_A \cdot U_{R_k} \cdot U_A^T - U_{L_{k+1}} \cdot U_{V_k} \cdot U_{L_{k+1}}^T$
\EndFor

\vspace{1mm}
\Statex \textbf{Output:}
\State Block encodings of conditional means \( \{ U_{\mu_1}, \dots, U_{\mu_T} \} \)
\State Block encodings of covariances \( \{ U_{R_1}, \dots, U_{R_T} \} \)
\end{algorithmic}
\end{algorithm}

\subsection{Unified Quantum Algorithm for LQG Control}

Algorithm \ref{alg:quantum-lqg} presents the complete quantum solution framework for the discrete-time finite-horizon output-feedback LQG control problem. The procedure integrates the two key components developed in the previous sections:

\begin{itemize}
    \item Backward Pass: Quantum Riccati recursion for computing the optimal LQR feedback gains \( \{K_0, \dots, K_{T-1}\} \) using backward Riccati recursion with block-encoded matrices and QSVT-based matrix inversion.
    
    \item Forward Pass: Quantum Kalman filtering for recursively estimating the conditional state \( \mu_k \), updating the error covariance \( R_k \), and computing the control input \( u_k = K_k \mu_k \) at each time step.
\end{itemize}

The backward pass is executed once, starting from the terminal cost matrix \( M_T \). The resulting feedback gains are then utilized in the forward pass to propagate the system state and apply the control policy at each time step k. 
In the presentation of Algorithm~\ref{alg:quantum-lqg}, for the sake of clarity, the computation of state estimates and control inputs is illustrated as two distinct steps, although they can be executed concurrently within a single forward pass.


\begin{algorithm}[t] 
\caption{Quantum Algorithm for Finite-Horizon Output-Feedback LQG Control}
\label{alg:quantum-lqg}
\begin{algorithmic}[1]
\Statex \textbf{Input:}
\State Time horizon \( T \) 
\State Block encodings \( U_A, U_B, U_C, U_M, U_N, U_S, U_\Sigma, U_\Gamma, U_\Upsilon \).
\State Initial estimate \( U_{\mu_0} \), initial covariance \( U_{R_0} \)
\State Block encodings of measurement stream \( \{ U_{y_1}, \dots, U_{y_T} \} \)
\Statex \textbf{Phase I: Backward Pass – LQR Gain Computation}
\State Compute block-encoded feedback gains \( \{ U_{K_0}, \dots, U_{K_{T-1}} \} \) using \textbf{Algorithm~\ref{alg:quantum-riccati}}.
\Statex \textbf{Phase II: Forward Pass – Kalman Filtering}
\State Run quantum Kalman filter using \textbf{Algorithm~\ref{alg:quantum-kalman}}, with inputs \( U_{\mu_0}, U_{R_0}, \{ U_{y_k} \}, \{ U_{K_k} \} \)
\State Obtain block-encoded state estimates \( \{ U_{\mu_1}, \dots, U_{\mu_T} \} \)
\Statex \textbf{Phase III: Control Computation}
\For{\( k = 0 \) to \( T-1 \)}
\State Compute classical control input: \( u_k \gets (U_{K_k} \cdot U_{\mu_k}) \)
\EndFor
\Statex \textbf{Output:}
\State Classical control inputs \( \{ u_0, \dots, u_{T-1} \} \)
\State Block-encoded conditional means \( \{ U_{\mu_1}, \dots, U_{\mu_T} \} \)
\end{algorithmic}
\end{algorithm}

Note that block-encoded matrices are not manipulated directly as data structures, but rather through unitaries acting on quantum states. Operations such as addition, multiplication, and inversion are implemented via compositions of quantum circuits. While we describe certain algorithmic steps using the language of matrix operations for clarity, all such operations are executed through corresponding quantum subroutines acting on block-encoded unitaries. This notation serves to concisely express the intended transformations, with the understanding that they are realized through quantum circuit implementations rather than direct manipulation of matrix entries.

\section{Complexity Analysis of the Quantum LQG Algorithm}
We now analyze the time complexity of the quantum algorithm proposed for solving the finite-horizon output-feedback LQG control problem. This analysis considers the core building blocks of the algorithm—block encoding, matrix arithmetic (addition and multiplication), and matrix inversion via QSVT \cite{gilyen2019quantum, chakraborty2018power}—and analyzes how these operations compose in the backward Riccati recursion and forward Kalman filtering stages.

\subsection{Time Complexity of Quantum Subroutines}
In this section, we evaluate the time complexity associated with each stage of the algorithm.
\paragraph{Block Encoding.}  
Let \( A \in \mathbb{C}^{n \times n} \) be a matrix stored in a quantum-accessible data structure. According to standard results, the matrix can be transformed into an \( (\alpha, a, \epsilon) \)-block encoding \( U_A \) in  \( \mathcal{O}(\mathrm{polylog}(n/\epsilon)) \) time. We denote this time as \( T_{U_A} \).
\paragraph{Matrix Addition.}  
Let \( U_A \) and \( U_B \) be block encodings of matrices \( A \) and \( B \), with parameters \( (\alpha, a, \epsilon) \) and \( (\beta, b, \epsilon') \), respectively. Using the linear combination of unitaries (LCU) method, we can construct a block encoding of \( A + B \) with parameters \( (\alpha + \beta, \max\{a, b\} + 1, \epsilon + \epsilon') \). The runtime for this addition is \( \mathcal{O}(T_{U_A} + T_{U_B}) \).
\paragraph{Matrix Multiplication.}  
Given the same encodings \( U_A \) and \( U_B \), a block encoding of the product \( AB \) can be constructed in time \( \mathcal{O}(T_{U_A} + T_{U_B}) \), following the multiplication rule for block-encoded matrices.
\paragraph{Matrix Inversion via QSVT.}  
To compute an approximate inverse of a Hermitian matrix \( A \), we use QSVT with a polynomial \( p(x) \approx 1/x \), which has degree \( d = \mathcal{O}(\kappa \log(1/\epsilon')) \), where \( \kappa \) is the condition number of \( A \). This requires \( \mathcal{O}(d) \) applications of block-encoded unitaries \( U_A \) and \( U_A^\dagger \), interleaved with controlled rotations. The overall time complexity of matrix inversion is thus \( \mathcal{O}(T_{U_A} \cdot d) = \mathcal{O}(T_{U_A} \cdot \kappa \log(1/\epsilon')) \).\\

\subsection{Time Complexity Analysis: Riccati Recursion for LQR}

We analyze the time complexity of the quantum backward pass used to compute the sequence of optimal state-feedback gains \( \{K_0, \dots, K_{T-1}\} \) via Riccati recursion, as described in Algorithm~\ref{alg:quantum-riccati}.

\paragraph{Block Encoding}  
At each time step \( k \), the algorithm operates on the system matrices \( A, B \), cost matrices \( M, N, S \), and the recursively updated Riccati matrix \( P_{k+1} \). All of these are assumed to be block-encoded using quantum-accessible data structures. The time to prepare each block encoding is denoted \( T_U = \mathcal{O}(\mathrm{polylog}(n/\epsilon)) \).

\paragraph{Matrix Addition and Multiplication}  
Each iteration involves computing intermediate matrices using block-encoded additions and multiplications for $U_1$ and $U_2$.
These operations require time \( \mathcal{O}(T_U) \) per matrix product or sum.

\paragraph{Matrix Inversion via QSVT}  
To compute the feedback gain matrix $K_k$, we apply QSVT to invert the block encoding of \( U_1 \). Assuming \( \kappa_R \) is the condition number of \( U_1 \), the inversion requires a polynomial of degree \( d = \mathcal{O}(\kappa_R \log(1/\epsilon'_R)) \), leading to a time complexity of \( \mathcal{O}(T_U \cdot \kappa_R \log(1/\epsilon'_R)) \).

\paragraph{Riccati Matrix Update}  
The matrix \( P_k \) updates consist of several multiplications and additions, with total cost per iteration also bounded by \( \mathcal{O}(T_U \cdot \kappa_R \log(1/\epsilon'_R)) \).\\

From the above, we can now state the following result.\\
\begin{proposition}[Quantum Riccati Recursion Time Complexity]
Let \( n \) be the system dimension, \( T \) the time horizon, and \( \epsilon, \epsilon'_R \) the allowable approximation errors for block encoding and QSVT-based inversion, respectively. Assume that all matrices involved in the Riccati recursion admit efficient block encodings and that the condition number \( \kappa_R \) of the inverted matrix \( B^\top P_{k+1} B + N \) is bounded.
Then, the total time complexity of the quantum Riccati recursion over a horizon of \( T \) steps is
\[
\mathcal{O}\left( T \cdot \mathrm{polylog}(n/\epsilon) \cdot \kappa_R \log(1/\epsilon'_R) \right).
\]
This represents an asymptotic improvement over the classical Riccati recursion, which typically has a per-iteration cost of \( \mathcal{O}(n^3) \) using dense matrix operations.
\end{proposition}

\subsection{Time Complexity Analysis: Kalman Filtering}
We analyze the time complexity of the quantum Kalman filtering procedure explained in Section \ref{alg:quantum-kalman}.
\paragraph{Block Encoding} Before performing quantum matrix operations, all required matrices must be stored in a quantum-accessible data structure. These include the system matrices \( A, C \), noise covariances \( \Sigma, \Gamma, \Upsilon \), and the initial state information \( \mu_0, R_0 \), along with dynamically updated variables such as \( R_k \), \( \mu_k \), and the Kalman gain \( L_k \). The time to block-encode each of these matrices is \( \mathcal{O}(T_U) \), where \( T_U = \mathrm{polylog}(n/\epsilon) \).
\paragraph{Innovation Covariance Construction}  
At each time step, the innovation covariance matrix is computed as \( V_k = \Gamma + C R_k C^\top \). The time complexity for computing this expression in the block-encoded form is \( \mathcal{O}(T_U) \), leveraging matrix multiplication and addition subroutines via the LCU framework.
\paragraph{Kalman Gain Computation}  
The Kalman gain \( L_k = W_k V_k^{-1} \) involves matrix multiplication and matrix inversion. The inversion of \( V_k \) via QSVT requires a polynomial of degree \( d = \mathcal{O}(\kappa_V \log(1/\epsilon'_V)) \), where \( \kappa_V \) is the condition number of \( V_k \). The total time complexity for the inversion is \( \mathcal{O}(T_U \cdot \kappa_V \log(1/\epsilon'_V)) \). The matrix product \( W_k V_k^{-1} \) can then be performed in time \( \mathcal{O}(T_U) \), yielding a total cost of \( \mathcal{O}(T_U \cdot \kappa \log(1/\epsilon'_V)) \) for computing \( L_k \).
\paragraph{State Estimate and Covariance Update}  
Each update requires a small constant number of block-encoded matrix multiplications and additions, each with time complexity \( \mathcal{O}(T_U) \). Thus, the full update step also incurs a cost of \( \mathcal{O}(T_U \cdot \kappa_V \log(1/\epsilon'_V)) \) per iteration. \\

\begin{proposition}[Quantum Kalman Filter Time Complexity]
Let \( n \) be the system dimension, \( T \) the time horizon, and \( \epsilon, \epsilon'_V \) the allowable approximation errors for block encoding and QSVT-based matrix inversion, respectively. Assume that all matrices involved in the Kalman filtering recursion admit efficient block encodings and that the condition number \( \kappa_V \) of the innovation covariance matrices is bounded.
Then, the total time complexity of the quantum Kalman filtering procedure over a horizon of \( T \) steps is
\[
\mathcal{O}\left( T \cdot \mathrm{polylog}(n/\epsilon) \cdot \kappa_V \log(1/\epsilon'_V) \right).
\]
This represents an asymptotic improvement over the classical Kalman filter, which incurs a cost of \( \mathcal{O}(n^3) \) per iteration under standard dense matrix operations.
\end{proposition}

Combining now the previous results, we obtain the LQG time complexity.\\

\begin{proposition}[Quantum LQG Time Complexity]
Let \( n \) be the system dimension, \( T \) the time horizon, and \( \epsilon, \epsilon'_R, \epsilon'_V \) the allowable approximation errors for block encoding and QSVT-based matrix inversion. Suppose that all system, cost, and covariance matrices admit efficient block encodings and that the matrices inverted during the Riccati and Kalman recursions have condition numbers \( \kappa_R \) and \( \kappa_V \), respectively.
Then, the total time complexity of the full quantum LQG algorithm is
\[
\mathcal{O}\left( T \cdot \mathrm{polylog}(n/\epsilon) \cdot \kappa \log(1/\epsilon') \right),
\]
where $\kappa=\max( \kappa_R, \kappa_V)$ and $\epsilon'=\min(\epsilon'_R,\epsilon'_V) $. \\
\end{proposition}

\section{Conclusion and Outlook}
In this work, we proposed a quantum algorithm for solving the discrete-time finite-horizon output-feedback LQG control problem. Our approach combines block-encoding-based quantum linear algebra with QSVT to implement both the backward Riccati recursion for optimal control synthesis and the forward Kalman filtering for optimal state estimation. By representing all system, cost, and noise matrices as block encodings, and leveraging efficient subroutines for matrix addition, multiplication, and inversion, we constructed a fully quantum pipeline for computing the control inputs of an LQG controller. 
We formally analyzed the time complexity of each component of the algorithm. Under standard assumptions on matrix condition numbers and encoding precision, the total time complexity of the quantum LQG algorithm scales polylogarithmically with the system dimension \( n \) and linearly with the time horizon \( T \), offering an asymptotic improvement over the classical \( \mathcal{O}(T n^3) \) complexity.

It is important to emphasize that the results in this paper are currently theoretical—relying on the availability of fault-tolerant quantum hardware and ideal quantum-accessible data structures. Nonetheless, the work provides a rigorous foundation for future exploration toward practical implementations of quantum control algorithms. It should also be acknowledged that each quantum subroutine used—such as QSVT-based matrix inversion and quantum matrix multiplication—introduces approximation errors. Although the algorithm is designed under standard assumptions that these errors remain bounded, a detailed error propagation analysis is necessary to rigorously confirm that the algorithm’s guarantees continue to hold under the cumulative effect of such quantum errors. 
Addressing this limitation is an important direction for future research, as it would enhance the practical feasibility and scalability of the proposed quantum control framework, particularly in settings involving dynamically changing system parameters, limited quantum memory, or imperfect block-encoding mechanisms.

To conclude, this work serves as a foundational step toward the design of scalable quantum algorithms for feedback control in high-dimensional systems. Future research directions include extending this framework to continuous-time systems, incorporating model uncertainty and robustness considerations, and developing end-to-end simulations and benchmarks using quantum hardware or quantum emulators.
\vspace*{0.4cm}

\bibliographystyle{IEEEtran}
\bibliography{IEEEabrv,ref}

\begin{thebibliography}{10}
\providecommand{\url}[1]{#1}
\csname url@samestyle\endcsname
\providecommand{\newblock}{\relax}
\providecommand{\bibinfo}[2]{#2}
\providecommand{\BIBentrySTDinterwordspacing}{\spaceskip=0pt\relax}
\providecommand{\BIBentryALTinterwordstretchfactor}{4}
\providecommand{\BIBentryALTinterwordspacing}{\spaceskip=\fontdimen2\font plus
\BIBentryALTinterwordstretchfactor\fontdimen3\font minus \fontdimen4\font\relax}
\providecommand{\BIBforeignlanguage}[2]{{%
\expandafter\ifx\csname l@#1\endcsname\relax
\typeout{** WARNING: IEEEtran.bst: No hyphenation pattern has been}%
\typeout{** loaded for the language `#1'. Using the pattern for}%
\typeout{** the default language instead.}%
\else
\language=\csname l@#1\endcsname
\fi
#2}}
\providecommand{\BIBdecl}{\relax}
\BIBdecl

\bibitem{anderson2007optimal}
B.~D. Anderson and J.~B. Moore, \emph{Optimal control: linear quadratic methods}.\hskip 1em plus 0.5em minus 0.4em\relax Courier Corporation, 2007.

\bibitem{whittle1990risk}
P.~Whittle, ``Risk-sensitive optimal control,'' \emph{Wiley}, 1990.

\bibitem{simon2006optimal}
D.~Simon, \emph{Optimal state estimation: Kalman, H infinity, and nonlinear approaches}.\hskip 1em plus 0.5em minus 0.4em\relax John Wiley \& Sons, 2006.

\bibitem{athans1971role}
M.~Athans, ``The role and use of the stochastic linear-quadratic-gaussian problem in control system design,'' \emph{IEEE transactions on automatic control}, vol.~16, no.~6, pp. 529--552, 1971.

\bibitem{shaiju2008formulas}
A.~Shaiju and I.~R. Petersen, ``Formulas for discrete time lqr, lqg, leqg and minimax lqg optimal control problems,'' \emph{IFAC Proceedings Volumes}, vol.~41, no.~2, pp. 8773--8778, 2008.

\bibitem{harrow2009quantum}
A.~W. Harrow, A.~Hassidim, and S.~Lloyd, ``Quantum algorithm for linear systems of equations,'' \emph{Physical review letters}, vol. 103, no.~15, p. 150502, 2009.

\bibitem{gilyen2019quantum}
A.~Gily{\'e}n, Y.~Su, G.~H. Low, and N.~Wiebe, ``Quantum singular value transformation and beyond: exponential improvements for quantum matrix arithmetics,'' in \emph{Proceedings of the 51st annual ACM SIGACT symposium on theory of computing}, 2019, pp. 193--204.

\bibitem{lin2022lecture}
L.~Lin, ``Lecture notes on quantum algorithms for scientific computation,'' \emph{arXiv preprint arXiv:2201.08309}, 2022.

\bibitem{sunderhauf2024block}
C.~S{\"u}nderhauf, E.~Campbell, and J.~Camps, ``Block-encoding structured matrices for data input in quantum computing,'' \emph{Quantum}, vol.~8, p. 1226, 2024.

\bibitem{clader2022quantum}
B.~D. Clader, A.~M. Dalzell, N.~Stamatopoulos, G.~Salton, M.~Berta, and W.~J. Zeng, ``Quantum resources required to block-encode a matrix of classical data,'' \emph{IEEE Transactions on Quantum Engineering}, vol.~3, pp. 1--23, 2022.

\bibitem{chakraborty2018power}
S.~Chakraborty, A.~Gily{\'e}n, and S.~Jeffery, ``The power of block-encoded matrix powers: improved regression techniques via faster hamiltonian simulation,'' \emph{arXiv preprint arXiv:1804.01973}, 2018.

\bibitem{shi2024quantum}
H.~Shi, G.~Zhang, and M.~Zhang, ``A quantum algorithm for the kalman filter using block encoding,'' \emph{arXiv preprint arXiv:2404.04554}, 2024.

\bibitem{clayton2024differentiable}
C.~Clayton, J.~Leng, G.~Yang, Y.-L. Qiao, M.~Lin, and X.~Wu, ``Differentiable quantum computing for large-scale linear control,'' \emph{Advances in Neural Information Processing Systems}, vol.~37, pp. 37\,176--37\,212, 2024.

\bibitem{krovi2024quantum}
H.~Krovi, ``Quantum algorithms to simulate quadratic classical hamiltonians and optimal control,'' \emph{arXiv preprint arXiv:2404.07303}, 2024.

\bibitem{camps2022fable}
D.~Camps and R.~Van~Beeumen, ``Fable: Fast approximate quantum circuits for block-encodings,'' in \emph{2022 IEEE International Conference on Quantum Computing and Engineering (QCE)}.\hskip 1em plus 0.5em minus 0.4em\relax IEEE, 2022, pp. 104--113.

\bibitem{camps2024explicit}
D.~Camps, L.~Lin, R.~Van~Beeumen, and C.~Yang, ``Explicit quantum circuits for block encodings of certain sparse matrices,'' \emph{SIAM Journal on Matrix Analysis and Applications}, vol.~45, no.~1, pp. 801--827, 2024.

\bibitem{martyn2021grand}
J.~M. Martyn, Z.~M. Rossi, A.~K. Tan, and I.~L. Chuang, ``Grand unification of quantum algorithms,'' \emph{PRX quantum}, vol.~2, no.~4, p. 040203, 2021.

\end{thebibliography}
\end{document}